\documentclass[11pt,superscriptaddress,aps,prd,preprint,nofootinbib]{revtex4}
\usepackage{amsmath}

\makeatletter
\usepackage[T1]{fontenc}
\usepackage{amsmath}
\usepackage{amssymb}
\usepackage{graphicx}
\usepackage{xcolor} 
\usepackage[normalem]{ulem}

\usepackage{slashed}

\newcommand{\bea}{\begin{eqnarray}}
\newcommand{\eea}{\end{eqnarray}}

\begin{document}

\title{G\"{o}del and G\"{o}del-type universes in Brans-Dicke theory}

\author{J. A. Agudelo}\email[]{jaar,alesandroferreira@fisica.ufmt.br}
\affiliation{Instituto de F\'{\i}sica, Universidade Federal de Mato Grosso,\\
78060-900, Cuiab\'{a}, Mato Grosso, Brazil}

\author{J. R. Nascimento}\email[]{jroberto,petrov,pporfirio@fisica.ufpb.br}
\affiliation{Departamento de F\'{\i}sica, Universidade Federal da Para\'{\i}ba\\
 Caixa Postal 5008, 58051-970, Jo\~ao Pessoa, Para\'{\i}ba, Brazil}

\author{A. Yu. Petrov}\email[]{jroberto,petrov,pporfirio@fisica.ufpb.br}
\affiliation{Departamento de F\'{\i}sica, Universidade Federal da Para\'{\i}ba\\
 Caixa Postal 5008, 58051-970, Jo\~ao Pessoa, Para\'{\i}ba, Brazil}

\author{P. J. Porf\'{i}rio}\email[]{jroberto,petrov,pporfirio@fisica.ufpb.br}
\affiliation{Departamento de F\'{\i}sica, Universidade Federal da Para\'{\i}ba\\
 Caixa Postal 5008, 58051-970, Jo\~ao Pessoa, Para\'{\i}ba, Brazil}

\author{A. F. Santos}\email[]{jaar,alesandroferreira@fisica.ufmt.br}
\affiliation{Instituto de F\'{\i}sica, Universidade Federal de Mato Grosso,\\
78060-900, Cuiab\'{a}, Mato Grosso, Brazil}
\affiliation{Department of Physics and Astronomy, University of Victoria,\\
3800 Finnerty Road Victoria, BC, Canada}

\begin{abstract}
In this paper, conditions for existence of G\"{o}del and G\"{o}del-type solutions in Brans-Dicke (BD) scalar-tensor theory and their main features are studied. 
The consistency of equations of motion, causality violation and existence of CTCs (closed time-like curves) are investigated. The role which cosmological constant and Mach principle play to achieve the consistency of this model is studied.

\end{abstract}

\maketitle

\section{Introduction}

Among the known exact solutions of Einstein field equations (EFEs)  gravity, the G\"{o}del and G\"{o}del-type metrics \cite{Godel1949, Novello1978, Hawking1975} play the special role. It was shown within the usual general relativity (GR) that these solutions describe rotating universes, allow for the existence of closed time-like curves (CTCs) and show that the Einstein theory of gravity is not completely compatible with Mach principle (MP)\footnote{According to this principle, the absolute acceleration does not exist, but the acceleration relative to distant cosmic matter distribution, while such matter determines inertial and geometrical properties of matter and space-time, respectively, can occur.} \cite{Adler1965, Reboucas1986, Dabrowski2003, Rosa2007, Buser2013}.

At the same time, the general relativity encounters several fundamental problems, such as its non-renormalizability at the quantum level and the need of explanation for the cosmic acceleration. To solve these problems, different alternative gravity theories were proposed (for a review on these theories, see \cite{Nojiri2006, Clifton2012}). Therefore, it is interesting to study the behaviour of the G\"{o}del and G\"{o}del-type solutions within these models, looking for the consistency of these metrics within such theories, and or their corresponding physical interpretations. Such studies, including discussion of problems of causality, the existence of CTCs and correspondence with GR in the respective limit, were performed through verification of the compatibility of resulting equations of motion in several gravity models including for example $f(R)$ gravity, Horava-Lifshitz gravity and bumblebee gravity \cite{Reboucas2009, Furtado2009, Santos2013, Fonseca2013, Nascimento2014, Santos2014}.

One more example of an alternative gravity model is the Brans-Dicke (BD) gravity which will be taken as the main subject of this paper. Here, we shall discuss the behaviour and main properties of G\"{o}del and G\"{o}del-type solutions in this theory, one of the first and well-known scalar-tensor theories, built up to be totally Machian and reducing to the GR in a limiting case \cite{Brans1961, Dehnen1971, Banerji1974}. To do this, we use the matter source composed by a perfect fluid and an electromagnetic field. Within our study, we verify the causality features of the possible solutions. Further, we find one completely causal solution corresponding to the empty space case. We note that earlier the $\omega\rightarrow \infty$ limit of BD theory was treated in \cite{Ferreira}.  

It is remarkable that the BD scalar field can be interpreted within different cosmological contexts, mainly within modelling the very early rapid expansion period known as inflation \cite{Linde1990, Artymowski2014}. Additionally, this field can be identified with the dilaton within the string theory context. Therefore, the BD model could be treated as a low energy limit of some unified and more general theory \cite{Fuji2004}.

However, it is known that the accelerated expansion cannot be described within a pure BD gravity. So, we must develop its possible extensions like the inclusion of cosmological constant (fixed or possessing different dependencies), scalar field potentials or functions of the scalar curvature \cite{Rador2007, DeFelice2010, Bisabr2012}. Thus, we review with the special attention the structure and solubility of resulting G\"{o}del and G\"{o}del-type field equations within BD theory, realistic cases and possible consequences. The role played by the cosmological constant and Mach principle as essential components for model coherence and compatibility is examined using the analogous and well-known results in GR. 

This work is structured as follows. In Section II, a brief review of fundamental ideas related to principles and properties of G\"{o}del and G\"{o}del-type universes are presented. Similarly, in Section III, the Brans-Dicke theory basics has been presented. In Section IV, the simple G\"{o}del universe in BD-$\Lambda$ model is studied. The Section V is devoted to study of the G\"{o}del-type universe in BD-$\Lambda$ model. In the Section VI, conclusions and remarks are presented. 


\section{G\"{o}del and G\"{o}del-type universes}

We start our paper with a brief review of main properties of G\"{o}del and G\"{o}del-type solutions of EFEs. 

\subsection{G\"{o}del case}

The simplest EFEs exact solution that allows CTCs is the G\"{o}del metric \cite{Godel1949}. This solution is compatible with incoherent matter distribution at rest and can be described by the line element looking like
\bea\label{2.1}
ds^2=a^2\bigl[(dx^0)^2-(dx^1)^2+\frac{e^{2 x^1}}{2}( dx^2)^2-(dx^3)^2+2e^{x^1}(dx^0 dx^2)\bigr],
\eea
where $a^2$ is a positive constant. This solution, for $a\neq 0$, is consistent only if the cosmological constant differs from zero. Therefore, considering an energy-momentum tensor of the pressureless relativistic fluid, $T^{\mu\nu}=\rho v^{\mu}v^{\nu}$, where $\rho$ is the matter density and $v^\mu$ is its $4$-velocity, it is simple to check that
\begin{equation}\label{2.2}
R_{\mu\nu}=-\frac{1}{a^2}v_\mu v_\nu,\qquad R=\frac{1}{a^2},
\end{equation}
so that the EFEs can be written in the form
\begin{equation}\label{2.3}
R_{\mu\nu}+(\Lambda-\frac{1}{2}R)g_{\mu\nu}=-8\pi GT_{\mu\nu},
\end{equation}
which implies that, in the system of units with $c=1$, the cosmological constant and matter density are
\begin{equation}\label{2.4}
\Lambda=-\frac{1}{2 a^2}, \qquad \rho=\frac{1}{8\pi G a^2}.
\end{equation}

It is worthwhile to mention some specific and important properties of this solution. We see that the energy-momentum tensor is the same as that one corresponding to the Einstein static universe, hence EFEs have two different solutions for the same matter content, which, from a purely Machian viewpoint, seems to be totally contradictory, since matter distribution should determine the space-time geometry uniquely \cite{Vishwakarma2015}. Thus, the G\"{o}del solution shows that GR has not satisfied Mach principle completely via its field equations. 


Additionally, this special solution describes a rotational cosmic behaviour, which can be seen clearly, comparing \eqref{2.1} with the metric corresponding to a flat space with cylindrical coordinates $r$, $\varphi$ and $z$, rotating around $z$-axis with angular velocity $\omega$, that is
\bea\label{2.5}
ds^2=(1-\frac{\omega^2 r^2}{c^2})c^{2}dt^2-dr^2-r^2d\varphi^2-dz^2-2\omega r^2 dtd\varphi,
\eea
which is analogous to the G\"{o}del solution (\ref{2.1}) through a natural correspondence $(x^0,x^1,x^2,x^3)\rightarrow(t,r,\varphi,z)$. 
Now, in order to describe quantitatively this rotational dynamics, one can introduce the following constants constructed on the base of the $4$-velocity:
\begin{equation}\label{2.6}
\Omega^\beta=c\frac{\epsilon^{\beta\mu\nu\gamma}}{\sqrt{-g}}a_{\mu\nu\gamma}, \quad a_{\mu\nu\gamma}=v_\mu \partial_\gamma v_\nu,
\end{equation}
where the 4-velocity is given by the vector 
\begin{equation}\label{2.8}
 v_\mu=(a,0,a e^{x^1},0), \quad v^\mu=(1/a,0,0,0),
\end{equation}
so, the $\Omega^\beta$ is the vorticity vector and $a_{\mu\nu\gamma}$ is a completely anti-symmetric tensor characterizing the orthogonality of geodesic trajectories within the G\"{o}del solution \cite{Godel2000}.

\subsection{G\"{o}del-type case}\label{gty}

It is a well known result that all G\"{o}del-type metrics, \textit{i.e.}, homogeneous space-times exhibiting vorticity, characterized by $\Omega$, and a given value of $m$  parameter\footnote{$-\infty<m^2=1/a^{4}<\infty$} \cite{Raychaudhuri1980,Reboucas1983}, can be rewritten in cylindrical coordinates as
\bea\label{2.9}
ds^2=dt^2+2H(r)dtd\varphi-G(r)d\varphi^2-dr^2-dz^2,
\eea
where the functions $G(r)$ and $H(r)$ must obey the relations
\begin{equation}
\begin{split}
&\frac{H^{'}(r)}{D(r)}=2\Omega,\\
&\frac{D^{''}(r)}{D(r)}=m^{2},
\end{split}
\label{HST}
\end{equation}
the prime denotes the derivative with respect $r$. The solution of Eqs.(\ref{HST}) can be divided in three different classes of G\"{o}del-type metrics in terms of $m^2$:

$\,\,$

i)\textit{hyperbolic class}: $m^2>0$,
\begin{equation}
\begin{split}
&H(r)=\frac{2\Omega}{m^2}[\cosh(mr)-1],\\
&D(r)=\frac{1}{m}\sinh(mr),\\
\end{split}
\end{equation}

$\,\,$

ii)\textit{trigonometric class}: $-\mu^2=m^2<0$, 
\begin{equation}
\begin{split}
&H(r)=\frac{2\Omega}{\mu^2}[1-\cos(\mu r)],\\
&D(r)=\frac{1}{\mu}\sin(\mu r),\\
\end{split}
\label{trigo}
\end{equation}

$\,\,$

iii)\textit{linear class}: $m^2=0$, 
\begin{equation}
\begin{split}
&H(r)=\Omega r^2,\\
&D(r)=r.\\
\end{split}
\label{linear}
\end{equation}

The case $m^2=2\Omega^2$ is a particular case of the hyperbolic class which corresponds to G\"{o}del solution \cite{Godel1949}.  It satisfies the relation $m^2=-2\Lambda=\kappa\rho=2\Omega^2$, where $\Lambda$ is the cosmological constant, $\rho$ is the matter density, $\Omega$ is the rotation and  $\kappa=8\pi G$, with $G$ being the gravitational constant.

An interesting aspect of G\"{o}del-type solutions is the possibility for existence of CTCs. The circle defined by $C=\lbrace(t,r,\theta,z); \, t=t_0, r=r_0, \theta \in [0, 2\pi], z=z_0\rbrace$, is a CTC if $G(r)$ becomes negative for a range of $r_c$ values ($r_1 <r_c<r_2$) \cite{Novello1978}, where $r_c$ is the critical radius, the minimal value of $r$ allowing for existence of CTCs. For the hyperbolic class ($m^2>0$) the critical radius is 
\begin{equation}
\label{2.11}
\sinh^{2}\left(\frac{mr_c}{2}\right)=\left(\frac{4\Omega^2}{m^2}-1\right)^{-1},
\end{equation}
such that it is valid on the range of parameters, $0<m^2<4\Omega^2$, and consequently there exists one non-causal region when $r>r_{c}$. On the other hand, the range, $m^2\geq 4\Omega^2$, does not present CTCs, i.e., the region is completely causal, for instance, the limiting case $m^2=4\Omega^2$ implies $r_{c}\rightarrow \infty$. The linear class ($m^2=0$) presents one non-causal region, $r>r_{c}$, such that the critical radius is given by $r_{c}=1/\Omega$. The trigonometric class ($m^2=-\mu^2<0$) presents an infinite sequence of alternating causal and non-causal regions \cite{Fonseca2013}. So is possible to have CTCs for all three classes.

Additionally, for the sake of the simplicity, we choose the basis\footnote{Indices $A$, $B$, $C$, $\ldots$ correspond to tangent space.}
\bea
ds^2=\eta_{AB}\theta^A \theta^B=\left(\theta^0\right)^2-\left(\theta^1\right)^2-\left(\theta^2\right)^2-\left(\theta^3\right)^2,
\eea
where the 1-forms $\theta^A=e^A_\alpha dx^\alpha$ are given by
\bea
\theta^0=dt+H(r)d\varphi, \quad\quad \theta^1=dr,\nonumber\\
\theta^2=D(r)d\varphi,\quad\quad \theta^3=dz.
\label{tetrad}
\eea
With this basis in the tangent space \cite{Reboucas1986}, it is possible to compute important quantities such as the Ricci scalar $R$ and the Einstein tensor $G_{AB}$, obtaining
\begin{equation}\label{2.12}
R=2(m^{2}-\Omega^{2}),
\end{equation}
and
\bea\label{2.13}
G_{00}&=&3\Omega^{2}-m^{2}\nonumber,\\
G_{11}&=&G_{22}=\Omega^{2},\\ 
G_{33}&=&m^{2}-\Omega^{2}.\nonumber
\eea
These results will be used in the next sections. 


\section{Brans-Dicke theory}

The Brans-Dicke (BD) theory is the first and the best motivated model introduced within the context of scalar-tensor gravity. It represents itself as a natural extension for the general relativity and was originally proposed to be totally compatible with Mach ideas and the weak equivalence principle (WEP) \cite{Brans1962,Fuji2004}. Within this theory, inertial masses of bodies and particles are treated as consequences of their interactions with some cosmic field rather then fundamental constants \cite{Weinberg1973}. 

Originally, Brans and Dicke suggested that the action of a new gravity theory should be similar to the Einstein-Hilbert action but including an additional non-minimal scalar field coupling:
\bea\label{3.1} 
 S=\int\sqrt{-g}\Bigl(\phi R-
 \frac{\omega}{\phi}\partial_{\mu}\phi\partial^{\mu}\phi+16\pi\mathcal{L}_{m}\Bigr)d^{4}x,
 \eea
where $R$ is the scalar curvature, $\phi$ is a scalar field treated as a some generalization of the gravitational constant ($\phi\propto G^{-1}$), which measures its scale locally. Further we will refer to it as to the BD field. Also, $\mathcal{L}_{m}$ is the matter Lagrangian which does not depend on $\phi$, so, $\partial_{\phi}\mathcal{L}_{m}=0$. Finally, the $\omega$ is a dimensionless constant representing itself as the unique free parameter in the theory. Varying this action with respect to $\phi$ and $g_{\mu\nu}$, we arrive at the original BD field equations looking like
\bea\label{3.2}
&&\frac{2\omega}{\phi}\square\phi-\frac{\omega}{\phi^{2}}\partial_{\mu}\phi\partial^{\mu}\phi+R=0,\\
\label{3.3}
&& R_{\mu\nu}-\frac{1}{2}g_{\mu\nu}R=\left(\frac{8\pi}{\phi}\right)T_{\mu\nu}+\frac{\omega}{\phi^{2}}\left(\partial_{\mu}\phi\partial_{\nu}\phi+
-\frac{1}{2}g_{\mu\nu}\partial_{\rho}\phi\partial^{\rho}\phi\right)
+\frac{1}{\phi}\left[\nabla_{\nu}(\partial_{\mu}\phi)-g_{\mu\nu}\square\phi\right],
\eea
with the covariant d'Alembertian operator acts on the BD field as
\begin{equation}\label{3.4} 
\square\phi=\nabla_{\mu}(\partial^{\mu}\phi)=
\frac{\partial_{\mu}\left(\sqrt{-g}\quad\partial^{\mu}\phi\right)}{\sqrt{-g}}.
\end{equation}
Multiplying the  Eq. (\ref{3.3}) by the inverse metric $g^{\mu\nu}$, we get
\begin{equation}\label{3.31} 
R=-\left(\frac{8\pi}{\phi}\right)T+\frac{\omega}{\phi^{2}}\partial_{\rho}\phi\partial^{\rho}\phi+\frac{3}{\phi}\square\phi,
\end{equation}
which we can combine with the Eq. (\ref{3.2}), obtaining
\begin{equation}\label{3.32} 
\square\phi=\left(\frac{8\pi}{3+2\omega}\right)T.
\end{equation}
This equation is evidently consistent with the Mach principle, because of the direct relationship between matter content characterized by $T$, and the BD field $\phi$ characterizing the inertial properties of the gravity. It is important to emphasize that, despite the matter and the BD field $\phi$ seem to be decoupled in the action of the theory, since they correspond to different contributions in the Lagrangian, they turn out to be strongly related because of this equation. Additionally, as a consequence of the fact that the matter Lagrangian does not depend on $\phi$, there is no possibility for spontaneous matter creation caused by BD field, since the energy-momentum tensor of matter obeys the $\phi$-independent equation
\begin{equation}\label{3.30} 
\nabla_{\nu}T^{\mu\nu}=0,
\end{equation}
satisfying hence the WEP.

Now, we plan to study the consistency of the G\"{o}del and G\"{o}del-type solutions within BD model. 


\section{G\"{o}del universe in Brans-Dicke gravity}

Consideration of the G\"{o}del solution within the BD gravity is equivalent to suggesting the possibility to have a non-stationary G\"{o}del solution, since the BD field $\phi$, should depend at least on the time $t$ \cite{Brans1961}. From now on we suppose that the scalar field $\phi$ depends either on the time $t$ or on $z$ coordinate. These dependencies have certain physical interpretations, for example, the $t$ dependence is motivated by cosmological reasons whereas the $z$ dependence -- by the axial symmetry characterizing the metric of G\"{o}del.

To study the G\"{o}del universe in BD gravity, one can rewrite the field equation (\ref{3.3}) as
\begin{equation}\label{5.1} 
R{^{\mu}_{\nu}}-\frac{1}{2}\delta{^{\mu}_{\nu}}R=\left(\frac{8\pi}{\phi}\right)T{^{\mu}_{\nu}}+\frac{\omega}{\phi^2}\left(\partial^{\mu}\phi\partial_{\nu}\phi-\frac{1}{2}\delta{^{\mu}_{\nu}}\partial_{\rho}\phi\partial^{\rho}\phi\right)+\phi^{-1}\left(\nabla_{\nu}\partial^{\mu}\phi-\delta{^{\mu}_{\nu}}\square\phi\right),
\end{equation}
and assume the energy-momentum tensor and 4-velocity of the matter to be given by
\begin{eqnarray}\label{5.3.1}
T_{\mu\nu}&=&\rho v_{\mu}v_{\nu},\nonumber\\
\label{5.3.2} 
v^{\mu}&=&(\frac{1}{a},0,0,0),\quad v_{\mu}=(a,0,ae^x,0).
\end{eqnarray}

As the simplest example, we assume the BD scalar to be only time dependent, $\phi=\phi(t)$, which corresponds to the cosmologically interesting situation (indeed, such a choice reflects the fact that the Universe is homogeneous and isotropic) one finds the components of the equation (\ref{5.1}) in the form 
\begin{align}
&(0,0):\quad\frac{1}{2a^2}=\left(\frac{8\pi}{\phi}\right)\rho-\frac{\omega}{2a^2}\left(\frac{\dot{\phi}}{\phi}\right)^{2}, \nonumber\\
&(i,i): -\frac{1}{2a^2}=\frac{\omega}{2a^2}\left(\frac{\dot{\phi}}{\phi}\right)^{2}+\frac{1}{a^2}\left(\frac{\ddot{\phi}}{\phi}\right), \label{5.6}\\
&(0,2): \quad\frac{1}{a^2}=\left(\frac{8\pi}{\phi}\right)\rho\nonumber,\\
&(1,2)=(2,1): \quad\left(\frac{\dot{\phi}}{\phi}\right)=0. \label{1.2}
\end{align}
where $i=1,2,3$.

However, this system turns out to be inconsistent except of the trivial case. Indeed, from the equation for the component (1,2) we obtain: 
\begin{equation}\label{5.7}
\phi(t)=C,
\end{equation}
where $C$ is an arbitrary constant, thus, the BD scalar turns out to be trivial. Therefore we conclude that for the case $\phi=\phi(t)$, the G\"{o}del metric in a pure BD model represents itself only as a trivial solution, with the BD scalar is reduced just to a constant, thus, the BD theory is reduced to the usual Einstein gravity. The natural question now is -- whether the BD gravity can be extended, and the G\"{o}del metric can be generalized, to achieve the consistency for the nontrivial BD scalar? To do it, we can consider the G\"{o}del-type metric originally proposed in \cite{Reboucas1983} and introduce the cosmological constant. In this context, we will consider another possibility for the $\phi$ field, that is, $\phi=\phi(z)$.


\section{G\"{o}del-type solution in BD-$\Lambda$ gravity}

The action of the BD-$\Lambda$ theory \citep{Tretyakova2011} can be written as
\bea
S=\frac{1}{16\pi}\int\sqrt{-g}\Bigl(\phi (R-2\Lambda)-
 \frac{\omega}{\phi}\partial_{\mu}\phi\partial^{\mu}\phi+16\pi\mathcal{L}_{m}\Bigr)d^{4}x.
\eea
For this study, we use the tangent space to make calculations simpler. Thus the field equations can be written as
\begin{equation}\label{4.19}
G{^{A}_{B}}-\delta{^{A}_{B}}\Lambda=
\left(\frac{8\pi}{\phi}\right)T{^{A}_{B}}+\frac{\omega}{\phi^2}\left(\partial^{A}\phi\partial_{B}\phi-\frac{1}{2}\delta{^{A}_{B}}\partial_{C}\phi\partial^{C}\phi
\right)+\phi^{-1}\left(\nabla_{B}\partial^{A}\phi
-\delta{^{A}_{B}}\square\phi\right),
\end{equation}
where
\begin{equation}
G_{AB}=e{^{\mu}_{(A)}}e{^{\nu}_{(B)}}G_{\mu\nu}, \quad T_{AB}=e{^{\mu}_{(A)}}e{^{\nu}_{(B)}}T_{\mu\nu},
\end{equation}
and
\begin{equation}
\eta_{AB}=e{^{\mu}_{(A)}}e{^{\nu}_{(B)}}g_{\mu\nu}, \quad \partial_{A}=e{^{\mu}_{(A)}}\partial_\mu, \quad \nabla_{B}=e{^{\nu}_{(B)}}\nabla_\nu.
\end{equation}

Now, we will add to our matter content an electromagnetic field aligned on $z$-axis and dependent of $z$, such a choice produces the following non-vanishing components of electromagnetic tensor in frame (\ref{tetrad})
\begin{equation}
  F_{(0)(3)}=-F_{(3)(0)}=E(z), \, F_{(1)(2)}=-F_{(2)(1)}=B(z),
  \label{rot}
  \end{equation}
with the solutions of the Maxwell equations are
\begin{equation}
\begin{split}
  &E(z)=E_{0}\cos[2\Omega(z-z_{0})],\\
  &B(z)=E_{0}\sin[2\Omega(z-z_{0})],
  \end{split}
\end{equation}
where $E_{0}$ is the amplitude of the electric and magnetic fields. Hence, the non-zero components of the energy-momentum tensor for the electromagnetic field are
  \begin{equation}
  T^{(\mathrm{ef})}_{(0)(0)}=T^{(\mathrm{ef})}_{(1)(1)}=T^{(\mathrm{ef})}_{(2)(2)}=\frac{E_{0}^{2}}{2}, \,\, T^{(\mathrm{ef})}_{(3)(3)}=-\frac{E_{0}^{2}}{2}.
  \end{equation}
   As a consequence, the new energy-momentum tensor is given by
   \begin{equation}
   T_{\mu\nu}=\rho v_{\mu}v_{\nu}+T^{(\mathrm{ef})}_{\mu\nu}.
   \end{equation}

    Next, we will find the solutions for the cases $\phi(t)$ and $\phi(z)$.

\subsection{\boldmath{$\phi=\phi(t)$}}\label{c}

In this case the d'Alembertian operator gets the form
\bea
\square\phi&=&\eta^{AB}\left[\partial_{B}(\partial_{A}\phi)-w{^{C}_{AB}}(\partial_{C}\phi)\right],\nonumber\\
\square\phi&=&\left(\frac{D^2-H^2}{D^2}\right)\ddot{\phi},
\eea
where $w{^{C}_{AB}}$ are the Ricci coefficients of rotation. Thus the diagonal components of the equations \eqref{4.19} are
\bea
&(0,0)&\quad 3\Omega^{2}-m^{2}-\Lambda=\left(\frac{8\pi}{\phi}\right)\rho+\left(\frac{4\pi}{\phi}\right)E^2_{0}+\frac{\omega}{2}\left(\frac{\dot{\phi}}{\phi}\right)^{2}+\frac{\omega}{2}\left(\frac{\dot{\phi}}{\phi}\right)^{2}\frac{H^2}{D^2}+\frac{\ddot{\phi}}{\phi}\frac{H^2}{D^2}, \nonumber\\
&(1,1)&\quad -\Omega^{2}-\Lambda=-\left(\frac{4\pi}{\phi}\right)E^2_{0}-\frac{\omega}{2}\left(\frac{\dot{\phi}}{\phi}\right)^{2}+\frac{\omega}{2}\left(\frac{\dot{\phi}}{\phi}\right)^{2}\frac{H^2}{D^2}-\frac{\ddot{\phi}}{\phi}\frac{D^2-H^2}{D^2}, \label{4.20}\\
&(2,2)&\quad -\Omega^{2}-\Lambda=-\left(\frac{4\pi}{\phi}\right)E^2_{0}-\frac{\omega}{2}\left(\frac{\dot{\phi}}{\phi}\right)^{2}-\frac{\omega}{2}\left(\frac{\dot{\phi}}{\phi}\right)^{2}\frac{H^2}{D^2}-\frac{\ddot{\phi}}{\phi}, \nonumber\\
&(3,3)&\quad \Omega^{2}-m^{2}-\Lambda=\left(\frac{4\pi}{\phi}\right)E^2_{0}-\frac{\omega}{2}\left(\frac{\dot{\phi}}{\phi}\right)^{2}+\frac{\omega}{2}\left(\frac{\dot{\phi}}{\phi}\right)^{2}\frac{H^2}{D^2}-\frac{\ddot{\phi}}{\phi}\frac{D^2-H^2}{D^2}, \nonumber
\eea
and the non-diagonal components are
\begin{align}
&(0,1)\quad \frac{HH'}{2D^2}\frac{\dot{\phi}}{\phi}=0,\nonumber\\
&(0,2)\quad \frac{H}{D}\left[\omega\left(\frac{\dot{\phi}}{\phi}\right)^{2}+\frac{\ddot{\phi}}{\phi}\right]=0,\\
&(1,2)\quad \frac{H'D-HD'}{D}\left[\omega\left(\frac{\dot{\phi}}{\phi}\right)^{2}+\frac{\ddot{\phi}}{\phi}\right]=0.\nonumber
\end{align}

A direct inspection of the component (0,1) implies that $\phi$ should be constant (we note that $H$ cannot be constant since it is fixed from the requirement of the space-time homogeneity Eq.(\ref{HST})). Therefore, in this case the G\"{o}del-type solutions in BD-$\Lambda$ gravity reduce to the GR solutions for one well-motivated matter whose solution is well known \cite{Reboucas1983}.

\subsection{\boldmath{$\phi=\phi(z)$}}\label{d}


In this case the d'Alembertian operator acts on $\phi$ as
\begin{equation}
\square\phi=-\phi'',
\end{equation}
and the non-zero components of the field equation are
\bea
\label{nonzero}
&(0,0)&\quad 3\Omega^{2}-m^{2}-\Lambda=\left(\frac{8\pi}{\phi}\right)\rho+\left(\frac{4\pi}{\phi}\right)E^2_{0}+\frac{\omega}{2}\left(\frac{\phi'}{\phi}\right)^{2}+\frac{\phi''}{\phi}, \nonumber\\
&(k,k)&\quad \Omega^{2}+\Lambda=\left(\frac{4\pi}{\phi}\right)E^2_{0}-\frac{\omega}{2}\left(\frac{\phi'}{\phi}\right)^{2}-\frac{\phi''}{\phi}, \\
&(3,3)&\quad \Omega^{2}-m^{2}-\Lambda=\left(\frac{4\pi}{\phi}\right)E^2_{0}-\frac{\omega}{2}\left(\frac{\phi'}{\phi}\right)^{2}.\nonumber
\eea

These field equations imply the relations
\begin{equation}\label{5.38}
4\Omega^{2}-m^{2}=\left(\frac{8\pi}{\phi}\right)(\rho+E_{0}^2), \quad m^2+2\Lambda=-\frac{\phi''}{\phi}.
\end{equation}
Discussing these equations \eqref{5.38}, it is possible to differ three cases, cf. \cite{Raychaudhuri1980, Reboucas1983}:

(i) If $\rho=0$ and $E_{0}=0$ the condition $4\Omega^{2}=m^{2}$ is found, since the solution belongs to the hyperbolic class and is completely causal. We note that this solution is consistent with the equation of motion of the scalar field
\bea
\label{scaleq}
-\phi''=\frac{1}{3+2\omega}(8\pi\rho+2\Lambda\phi).
\eea
(the trace of the energy-momentum tensor is $T=\rho$ (it does not depend of $E_{0}$) which in this case gives zero), so, the equation for $\phi$ yields the exponential solution $\phi(z)=C_1e^{kz}+C_2e^{-kz}$, with $k=\sqrt{-\frac{2\Lambda}{3+2\omega}}$, this form of the solution is valid when $\Lambda<0$. If $\Lambda>0$ we get $\phi(z)=C_{3}\cos(k^{'}z)+C_{4}\sin(k^{'}z)$, where $k=ik^{'}$. Using Eqs. (\ref{5.38}-\ref{scaleq}) the $m$ parameter is related with $k$ through the relation
\bea
m^2+2\Lambda=\frac{2\Lambda}{3+2\omega}\quad\quad \Longrightarrow \quad\quad \frac{m^2}{4}+\Lambda=-k^2\left(1+\frac{\omega}{2}\right).
\eea
We note that $\Lambda$ plays a important role since the parameters of the metric $m^2$ and $\Omega^2$, as well as the field $\phi$, can be written in terms of it. Then this case (that is, the vacuum solution) represents one completely causal solution of the G\"{o}del-type universe in the BD-$\Lambda$ formalism. The same solution ($m^2=4\Omega^2$) has been obtained in GR-$\Lambda$ context for the massless scalar field as the only matter source \cite{Reboucas1983}. In addition, in the limit $\omega\rightarrow \infty$ we can show to similarity among BD and GR theory (it is expected that in this limit the BD field equations reduce to GR field equations for the same energy-momentum tensor, for more discussions on this issue see \cite{Romero}-\cite{Capozziello}). By taking this limit we get $\phi(z)\approx\phi_{0}(1 \pm kz)$, where $C_1=C_2=\phi_{0}=1/G$.

Using this limit and the vierbein, i.e, $\partial_{A}=e^{\,\,\,\mu}_{_{A}}\partial_{\mu}, \,\, \partial^{A}=\eta^{AB}\partial_{B}$, one can rewrite eq. (\ref{4.19}) as 
\bea
\label{omega}
G_{AB}=\Lambda\eta_{AB}-\frac{1}{2}\Lambda\sigma_{AB}+O(1/\sqrt{\omega}),
\eea
where $\sigma_{(3)(3)}=-3$ and $\sigma_{AB}=\eta_{AB}$, with $A,B\neq 3$. Therefore, in this case, the solution for $\omega\rightarrow \infty$ does not recover the vacuum Einstein field equations, as shown in \cite{Faraoni} when the trace of the energy-momentum tensor vanishes the BD theory. However, we can interpret the term $\frac{1}{2}\Lambda\sigma_{AB}$ in Eq. (\ref{omega}), as one contribution to the energy-momentum tensor and recover the same completely causal solution obtained in \cite{Reboucas1983} when $\omega\rightarrow \infty$.

In this way, we conclude that the vacuum-solution of BD-$\Lambda$ field equations for G\"{o}del-type metrics is completely causal and, in the limit $\omega\rightarrow\infty$ is similar the GR-$\Lambda$ with a matter source specific, as for instance the scalar field ($\phi(z) \propto z$) used in work \cite{Reboucas1983}.

(ii) If $\rho>0$, it is necessary to require $\frac{\rho}{\phi}=const$, in order to get solutions consistent with the Eqs. (\ref{5.38})-(\ref{scaleq}). In addition, the solutions are restricted by the interval, $m^2<4\omega^2$, thus it is possible to carry out the following analysis of the solutions (\ref{5.38}):
\begin{itemize}
\item $0< m^2<4\Omega^2$ - solutions of the hyperbolic class,  there is one noncausal region for a given $r>r_c$ given by Eq. (\ref{2.11}) ;
\item $m^2=0$ - solutions of the linear class, since there is one noncausal region for a given $r>r_c$ given by $r_{c}=1/\omega$ ;
\item $m^2=-\mu^2<0$ - solutions of the trigonometric class, thus there is an infinite number of alternating causal and noncausal regions.
\end{itemize} 

However, in this case, if $\rho$ is constant, one should have $\phi=const$ as well, and the situation becomes trivial reducing to the usual Einstein gravity.
The possible nontrivial solutions can look like $\rho(z)=C_{1} \phi(z)=C_{1}/G_{eff}(z)$, such  that the decreasing of the effective gravitacional coupling, $G_{eff}(z)$, implies the growing of the density($\rho$) and reciprocally, for $C_{1}>0$. However, for the linear class ($m^2=0$) we have nontrivial solution. We choose
 $\rho=C_1\cos{kz}$, $\phi=C_2\cos{kz}$, with $E_0=0$. In this case, the equations (\ref{nonzero}) become purely algebraic ones:
\bea
\label{nonzero1}
\Omega^2=2\pi\frac{C_1}{C_2}, \\
\Lambda=\frac{2\pi\frac{C_1}{C_2}}{1+\omega}. 
\eea
From these equations, one can find $\Omega^2$ and $\Lambda$ as functions of the parameters $\omega,C_1/C_2$. It is clear that the equation (\ref{scaleq}) is also consistent with these solutions yielding the relation $k^2C_2=\frac{1}{3+2\omega}(8\pi C_1+2\Lambda C_2)$. 

(iii) If $\rho<0$, the condition $4\Omega^2<m^2$ takes place when $\rho<-E^2_{0}$. This condition implies that there is no CTC in the corresponding G\"{o}del-type space-time. However, again, a constant $\rho$ implies a constant $\phi$ as well, which reduces the situation to the usual Einstein gravity, with this solution itself is excluded since it corresponds to $m^2\leq 2\Omega^2$ (that is, just the result following from our equations at $\phi=const$), which is incompatible with our condition $4\Omega^2<m^2$. So, this situation is inconsistent.

At the same time, one can notice that from equation $m^2+2\Lambda=-\frac{\phi''}{\phi}$ the scalar field is found as
$$\phi(z)=C\cos\gamma z+D\sin\gamma z,$$ where $\gamma^2=2\Lambda+m^2$. Also, one can see that the original G\"{o}del universe, $m^2=-2\Lambda$, is only possible if the scalar field is constant.


\section{Conclusions}

We discussed the G\"{o}del-type solutions within the context of the BD gravity.  In our study, the consistency of G\"{o}del solution within BD gravitational formalism has been reviewed, and we showed the importance of the non-zero cosmological constant $\Lambda$ in order to have a non-trivial solution, describing the expected values of different parameters analogous to the GR. The field equations of the BD-$\Lambda$ formalism were solved for the two cases $\phi(t)$ and $\phi(z)$ both for G\"{o}del solution and G\"{o}del-type solution. For the G\"{o}del solution in BD-$\Lambda$ formalism the consistency and correspondence with GR were verified. Afterwards, the G\"{o}del-type solution with $\phi(t)$ was considered, and a condition allowing to reduce this solution to the original G\"{o}del metric was determined by the system of equations. When the scalar field depends only on $z$, there are two possibilities depending on the sign of the matter density: (i) empty causal G\"{o}del-type universe, which corresponds to the exact solution with $m^2=4\omega^2$, such a solution has been obtained with the requirement of $\Lambda\neq 0$. Additionally, we verified that, for the constant density case (which is a more usual situation within the cosmological studies since it reflects the large-scale homogeneity and isotropy of the space), in the limit $\omega\rightarrow \infty$, this solution reduces to GR with one massless scalar field, with some implications associated to cosmological constant; (ii) both causal and non-causal regions are allowed. Therefore, our study also shows that the idea of BD theory as a totally Machian theory should be revisited and discussed in more details.


\acknowledgments
This work was partially supported by Coordena\c{c}\~ao de Aperfei\c{c}oamento de Pessoal de N\'{\i}vel Superior (CAPES),  and Conselho Nacional de Desenvolvimento Cient\'{\i}fico e Tecnol\'{o}gico (CNPq). J. A. Agudelo would like to thank to all colleagues of Instituto de F\'{\i}sica in UFMT-Cuiab\'{a} by the shared knowledge and time. A. F. S. has been supported by the CNPq projects 476166/2013-6 and 201273/2015-2. The work by A. Yu. P. has been supported by the CNPq project 303783/2015-0.


\end{document}